\documentclass[aps, reprint, superscriptaddress,nofootinbib, showpacs,preprintnumbers, floatfix]{revtex4-1}
\usepackage{amsmath, latexsym, amssymb, hyperref, graphicx, color, multirow, slashed}
\usepackage[table,xcdraw,dvipsnames]{xcolor}
\usepackage{tabularx}
\usepackage[T1]{fontenc}
\usepackage[capitalize]{cleveref}
\definecolor{nicered}{rgb}{.7,.1,.1}
\definecolor{nicegreen}{rgb}{.2,.7,.1}
\definecolor{lightgreen}{rgb}{.6,.9,.5}
\definecolor{darkblue}{rgb}{0,0,.5}
\definecolor{el}{rgb}{.9, .8, .7}
\definecolor{mu}{rgb}{.8, .7, .8}
\hypersetup{colorlinks, citecolor=nicegreen,linkcolor=nicered, urlcolor=darkblue}
\usepackage{caption}
\usepackage{subcaption}
\usepackage[normalem]{ulem}

\definecolor{nicered}{rgb}{0.7,0.1,0.1}
\definecolor{nicegreen}{rgb}{0.1,0.5,0.1}
\definecolor{red}{rgb}{1.0, 0, 0}

\newcommand\RR{\mathbb R}
\renewcommand{\L}{{\cal L}}
\renewcommand{\P}{{\cal P}}
\newcommand{\C}{{\cal C}}
\newcommand{\T}{{\cal T}}

\newcommand{\SU}{{\rm SU}}
\newcommand{\SO}{{\rm SO}}

\newcommand{\U}{{\rm U}}
\newcommand\e{\mathrm e}

\newcommand\g{\gamma}
\renewcommand\a{\alpha}
\newcommand\s{\sigma}

\newcommand\w{\omega}
\newcommand\1{{\mathbf 1}}
\newcommand\0{{\mathbf 0}}

\renewcommand{\[}{\left[}

\newcommand\PM[1]{\begin{pmatrix}#1\end{pmatrix}}
\newcommand{\bdm}{\begin{displaymath}}
\newcommand{\edm}{\end{displaymath}}
\newcommand{\bea}{\begin{eqnarray}}
\newcommand{\eea}{\end{eqnarray}}

\newcommand{\nn}{\nonumber}

\renewcommand\P{\mathcal P}
\renewcommand{\a}{\alpha}

\renewcommand{\t}{\theta}

\definecolor{nicered}{rgb}{0.7,0.1,0.1}
\definecolor{nicegreen}{rgb}{0.1,0.5,0.1}
\definecolor{red}{rgb}{1.0, 0, 0}
\definecolor{niceblue}{rgb}{0,0,0.8}
\definecolor{red}{rgb}{1.0, 0, 0}
\hypersetup{colorlinks,citecolor= nicegreen,linkcolor= nicered,urlcolor=nicered}

\def\diag{\mbox{diag}\,}

\def\gsim{\raise0.3ex\hbox{$\;>$\kern-0.75em\raise-1.1ex\hbox{$\sim\;$}}}
\def\lsim{\raise0.3ex\hbox{$\;<$\kern-0.75em\raise-1.1ex\hbox{$\sim\;$}}}

\def\mb[#1]{\mathbf{#1}}

\renewcommand{\bar}{\overline}

\definecolor{LightCyan}{rgb}{0.88,1,1}
\definecolor{piggypink}{rgb}{0.99, 0.87, 0.9}
\definecolor{applegreen}{rgb}{0.55, 0.71, 0.0}
\definecolor{darkpastelgreen}{rgb}{0.01, 0.75, 0.24}
\definecolor{green-yellow}{rgb}{0.68, 1.0, 0.18}

\newcommand{\be}{\begin{equation}}
\newcommand{\ee}{\end{equation}}
\newcommand{\beq}{\begin{equation}}
\newcommand{\eeq}{\end{equation}}
\newcommand{\beqa}{\begin{eqnarray}}
\newcommand{\eeqa}{\end{eqnarray}}
\newcommand{\ba}{\begin{array}}
\newcommand{\ea}{\end{array}}

\newcommand{\Sec}[1]{ \medskip \noindent {\sl \bfseries #1}}

\begin{document}


\title{Parity from gauge symmetry}
\author{Alessio Maiezza}
\email{alessiomaiezza@gmail.com}
\affiliation{\normalsize \it Rudjer Bo\v skovi\'c Institute, Bijeni\v cka cesta 54, 10000, Zagreb, Croatia,}

\author{Fabrizio Nesti}
\email{fabrizio.nesti@aquila.infn.it}
\affiliation{\normalsize \it
  Dipartimento di Scienze Fisiche e Chimiche, Universit\`a degli Studi dell'Aquila, via Vetoio, I-67100, L'Aquila, Italy}
\affiliation{\normalsize \it INFN, Laboratori Nazionali del Gran Sasso, I-67100 Assergi (AQ), Italy}

\begin{abstract}
\noindent
We argue that Left-Right parity symmetry $\P$ can arise as a discrete remnant of a unified gauge
symmetry.  The high-energy unification necessarily includes the gauging of the Lorentz symmetry,
bringing into the game gravitational interactions, and leading to a gravi-GUT scheme. Parity emerges
unbroken below the Planck scale, and can be broken spontaneously at lower energies making contact
with the Standard Model.  This framework motivates the spontaneous origin of parity violation as in
Left-Right symmetric theories with $\P$. The possible unifying gauge groups are identified as
$\SO(1,7)$ for gravitational and weak interactions, or $\SO(7,7)$ for a complete unification.
\end{abstract}

\maketitle

\Sec{Introduction.}  The chiral asymmetry of weak interactions has been discussed since the seminal work of Lee and Yang on parity violation~\cite{Lee:1956qn}, where the possibility of its restoration was advocated in terms of mirror particles.  Rather than duplicating the matter
spectrum, a different restoration of parity is achieved in the popular Left-Right symmetric models (LRSM) by extending the weak gauge group, as
$\SU_L(2)\times \SU_R(2) \times \U_{B-L}(1)$~\cite{Senjanovic:1978ev,Pati:1974yy,Mohapatra:1974hk,Mohapatra:1974gc,Senjanovic:1975rk}, see~\cite{Senjanovic:2011zz} for a review.

Parity restoration demands a discrete symmetry exchanging left with right, which can be realized either as a left-right parity $\P$ or as a left-right charge conjugation $\C$~\cite{Maiezza:2010ic}.
The latter has a natural UV protection in $\SO(10)$ Grand Unified (GUT) models, as $\C$ can be found among the gauge generators~\cite{Kibble:1982dd,Aulakh:2002zr}.  The former, on the other hand, is the original and preferred choice if one aims for a true parity-conserving theory at high energy but lacks a UV completion.  In this Letter, we propose a possible solution to this long-standing problem.

In analogy with $\C$ within $\SO(10)$, one would arrange $\P$ as a generator of a unified gauge
group, such that the discrete symmetry at low energy can be interpreted as a remnant of a continuous
one.  However, this approach for $\P$ is hampered by the fact that a continuous rotation mixing
chiralities does not commute with the Lorentz symmetry. Probably, this is the main obstacle to
formulating a theory of $\P$.  A possible approach, rooted in the idea of Kaluza-Klein
theories~\cite{Witten:1981me}, considers parity as part of the 5D Lorentz and coordinate
transformations so that it can be obtained as a discrete remnant symmetry in 4D, an idea explicitly
considered in e.g.\ Ref.~\cite{Choi:1992xp}. The price to pay in this approach is the dependence on
the unknown dynamics of dimensional reduction, in addition to issues in anomaly
matching~\cite{Wetterich:1983bg,Alvarez-Gaume:1983ihn}.

However, since the crucial point is the non-commutation of parity with the Lorentz symmetry, extra
dimensions are not strictly necessary: an effective and more economical approach is to promote both
as part of unified \emph{internal} gauge symmetry.  At high energy, this internal symmetry is
completely disentangled from space-time diffeomorphisms, while they are soldered below a breaking
scale, where a standard Lorentzian physics emerges.  This framework necessarily brings into the game
also gravity, in Cartan formulation.

In addition, since in the real world left and right fermions have different weak charges, one shall
mix parity not just with Lorentz, but also with the Standard Model (SM) gauge symmetries, leading to
a unified group whose gauge fields include the gravitational connection along with standard gauge
fields. This is the approach put forward in~\cite{Cahill:1982zf,Percacci:1984ai} and implemented as
gravi-weak or gravi-GUT setups~\cite{Nesti:2009kk,Nesti:2007ka,Nesti:2007jz}.  The unifying group is
spontaneously broken by a vacuum expectation value (VEV) of an extended vierbein field at the Planck
scale, leading to an unbroken gauge subgroup plus the residual \emph{global} Lorentz symmetry, as we
shall review below.\footnote{For unifications involving extra dimensions, 
see~\cite{Trayling:2001kd,Maraner:2003sq,Krasnov:2018txw,Krasnov:2021jjp,Vaibhav:2021xib}.}

It is thus interesting to uncover the role of parity in this framework. In the present Letter, we
investigate the viability of this approach from the point of view of symmetries and show that
realizing parity will lead us to select a gravi-weak scenario based on the $\SO(1,7)$ gauge
symmetry, and a complete unification for $\SO(7,7)$.

We will conceptually decompose Left-Right parity $\P$ in two operations: the inversion of space
$I_s$ and the internal action on fields, which we name $P$.  These two operations can be
disentangled at high energy, allowing for the gauging of the internal part $P$, i.e.\ its
protection.  At low scale, they are soldered and give rise to $\P$.  Thus, $\P$ is automatic and
protected by the gauging if the theory is assumed to respect basic spatial inversion $I_s$. If, on
the other hand, we allow for inversion-violating terms at high energy, still internal parity is
gauged and just a few $\P$-breaking terms are allowed to emerge, notably the topological QCD
$F\tilde F$ term, unified with its gravitational analogue $R\tilde R$.

Finally, we will argue that unlike the usual breaking of gauge to discrete symmetries, the proposed framework does not lead to cosmic strings.

\smallskip

We will discuss first the internal symmetry part and later the breaking which connects with space-time, and finally the implications of this idea.

\Sec{Making parity action continuous.}  Looking at the action of parity on fields, and ignoring
for the moment the weak isospin, we denote, using a Weyl basis in 1+3 dimensions,
\be
\Psi=\PM{\psi_L\\\psi_R} \,, \quad P=\g_0\equiv\PM{\0&\1\\\1&\0}\equiv\1\otimes\s_1\,.
\ee
As usual, parity swaps fermions as
\be
\psi_L\leftrightarrow \psi_R    \quad \text{or}\quad   \Psi \to P\, \Psi\,.
\ee

Now, the discrete $P$ can be enlarged to a continuous symmetry $U(1)_P$ in the
$(\psi_L, \psi_R)$ space, as follows:
\be
U(\a) = \e^{i\frac\a2 X (P-1)} = \e^{-i\frac\a2 X} \left(\cos\frac\a2 + i XP \sin\frac\a2\right),
\label{eq:Pcont}
\ee
where $X$ is any matrix which commutes with $P$ and has $X^2=1$. As readily checked, parity is a
rotation by $\pi$:
\be
U(\pi)=P\,,\qquad U(0)=U(2\pi)=1\,.
\label{eq:UP}
\ee
While $X=1$ is a possibility, the other and more interesting choice, to appear in the
following, is $X=n_i\s_i\otimes\1$ with $n^2=1$, for instance $X=\s_3\otimes\1$.

\smallskip

\Sec{Unifying with Lorentz group.}  To promote the above $U(1)_P$ to a gauge symmetry, one is
clearly faced with the fact that parity does not commute with the action of the Lorentz group, in particular, it commutes with angular momentum but not with boosts.

Thus Lorentz has to be included. The simplest and illustrative example is provided by enlarging the
Lorentz group to $\SO(1,4)$, which contains 3D parity and has precisely $\Psi$ as its non-chiral
4-dimensional spinor representation. Labeling the internal directions as $0,\ldots,4$, the new
internal spacelike direction 4 requires four new generators: three rotations $R_i$ in the $i$-4
planes, and one boost $K_4$. One can write
\bea
L_i = \frac{\s_i\otimes\1}2\,,&\quad&  K_i = \frac{i\s_i\otimes\s_3}2\,,\nn\\
R_i = \frac{\s_i\otimes\s_1}2\,,&\quad&  K_4=\frac{i\1\otimes\s_2}{2}\,.
\eea

Now, one notes that the angular momentum $L_i$ commutes with $P$ and that $R_i = L_i P$. Comparing
then with (\ref{eq:Pcont}) we see that $U(1)_P$ is precisely generated by $L_i-R_i$, for whichever
$i=1,2,3$. Thus, parity is interpreted geometrically in terms of the new spatial direction:
choosing for instance $i=3$, parity $P=U(\pi)$ consists of two simultaneous rotations by $\pi$: one
among 3-4 generated by $XP=R_3$, and one among 1-2 generated by $L_3$.  After these $\pi$ rotations,
effectively the 123 directions are reversed (and so is 4).  It leads thus to spinor exchange
(\ref{eq:UP}) plus internal spatial reflection. Different $i$ imply different rotation planes, but
after the $\pi$ rotation the final effect is the same.

\smallskip

This first example misses the fact that $\psi_L$ and $\psi_R$ in the SM belong to different gauge
multiplets. In particular, they have different weak and hypercharge representations or, in the
language of LR symmetry, they transform under different $\SU(2)_{L,R}$ groups. Therefore, a
realistic example must involve at least the weak interactions, as we shall discuss now.

\Sec{\SO(1,7) example.}  \SO(1,7) has a Majorana representation of real dimension 16, that can be
mapped into complex dimension 8 and that, under the decomposition
$\SO(1,7)\to\SO(1,3)\times\SO(0,4)$, leads precisely to the required pattern where $\psi_L$ and
$\psi_R$ transform as doublets under the $\SU(2)_L$, $\SU(2)_R$ components of $\SO(4)$,
\be
\mathbf{16}_\RR\equiv\mathbf{8}_s \to (\mathbf{2}_L,\1_R, \mathbf{2}_l) \oplus (\1_L,\mathbf{2}_R, \mathbf{2}_r)\,.
\ee
Here the first two slots refer to SU(2)$_{L,R}$ and the last to Lorentz, namely $2_l$, $2_r$ for
left and right Weyl spinors. It is convenient to spell out the $\SO(1,7)$ generators acting on
$\mathbf{8}_s$ as $(2/i)\Sigma_{M,N}$:
%
\def\ot{\mathop{\scriptstyle\otimes}}
{%
  \smallskip
    \small
$$
 %
\left(
\begin{array}{cc|cc}
 0& i\s_i \ot  \1 \ot\s_3  &\! i \1 \ot \1 \ot\s_2  & i \1 \ot \s_b\ot\s_1 \\[1ex]
  \!  -i\s_j \ot \1 \ot\s_3 &   \epsilon_{ijk}  \s_k\ot \1\ot\1              &\!   -\s_j \ot \1\ot\s_1  & \s_b\ot \s_j\ot \s_2 \\[2ex]
  \hline
  && \\[-1ex]
  -i \1 \ot  \1 \ot\s_2  & \fbox{$\s_i \ot \1\ot\s_1$}  & 0   &  -\1 \ot\s_b\ot \s_3 \\[1ex]
  -i \s_b\ot \1 \ot \s_1   & -\s_i\ot \s_b\ot \s_2    &  \1 \ot\s_a\ot \s_3&   \,  \epsilon_{abc}  \1 \ot\s_c\ot \1
\end{array}
\right)
$$
}%

\noindent
where, out of the internal directions $M,N=0,\ldots,7$, the first four (0123) are along Lorentz. Thus
the upper-left block represents the $\SO(1,3)$ generators and the lower-right the $\SO(4)$ ones. In the
respective spaces we denoted $i,j$ or $a,b$ as indices from 1 to 3, thus matching $M,N=1,2,3\to i,j=1,2,3$
and $M,N=5,6,7\to a,b=1,2,3$.

The three boxed generators 123-4 correspond precisely to $R_i\sim\s_i\ot\1\ot\s_1$, while just above
one finds the rotations $L_i\sim\s_i\ot\1\ot\1$.  As before, we find that for any $i$ the
combination $R_i-L_i$ generates a $\U(1)_P$ and a rotation by $\pi$ generates
$P=U(\pi)=\1\ot\1\ot\s_1$. Here the novelty is that this time, under $P$, the swapping of left and
right spinors is accompanied by the swapping of the Left and Right weak groups, as required in a
realistic model.\footnote{The further nine generators below the boxed ones also lead to $P$, modulo
  a $\SO(4)$ gauge rotation.}

\smallskip

\Sec{Symmetric phase and breaking.}  We can turn now to describe a mechanism of symmetry breaking
which preserves parity as a discrete remnant of the original continuous gauge group.  The task is
complicated by the unification of local Lorentz with the other gauge forces, and by the fact that
spacetime symmetries must be involved in such a way that at low energy $\P$ includes the spacetime inversion.

As anticipated, in the first-order approach the Lorentz symmetry is disentangled from spacetime
transformations (diffeomorphisms) and treated as an internal gauge symmetry, further extended to
include other interactions.  

The framework is based on the first-order (Cartan) formulation of Gravity (see
e.g.~\cite{Castellani:1991et}) where the gauge field of the Lorentz group is a spin connection
$\w^m_{\mu\,n}$ and the vierbein field $e_\mu^m$ transforms as a vector under the local Lorentz
group (index $m=0,\ldots,3$).
A background value (VEV) of the vierbein is needed for a sensible low energy spacetime metric: for a
standard Minkowski background it is $\bar e_\mu^m=M_{pl}\,\delta_\mu^m$, with $M_{pl}$ the
Planck mass.  This VEV, regarded as the choice of a fixed (unitary) gauge, breaks both the local
Lorentz group and the diffeomorphism invariance.  It nevertheless leaves unbroken a joint
\emph{global} Lorentz symmetry, realized when Lorentz and diff transformations on $\mu$ and $m$ are
matched.  This is the global Lorentz invariance of the Minkowski background that we experience at
low energy.  The counting of degrees of freedom confirms that of the 16 independent fields in
$e_\mu^m$, 6 fields correspond to the gauge modes of local Lorentz transformations and are set to
zero with the gauge fixing, or ``eaten'' by the spin connection, which can be shown to acquire a
mass of the order of the Planck scale $M_{pl}$.  The other 10 degrees of freedom become propagating
and carry the standard graviton.  In this formulation, the vierbein acts as a Higgs field for
the breaking of the local Lorentz group to a global symmetry.

It is worth stressing again that in the symmetric phase the internal (gauge) Lorentz transformations
are disentangled from the spacetime (diff) ones, while in the broken phase they are glued.
Accordingly, spinors are originally scalars under spacetime transformations and just transform under
internal local Lorentz.  Only in the broken phase they become spinors also of spacetime transformations,
being these glued to Lorentz.  As an example, their fermionic kinetic term arises from a
symmetric-phase lagrangian written geometrically as
\be
\L_{\psi\, kin}=\bar \psi\gamma^m D \psi \wedge e^n \wedge e^r \wedge e^s \epsilon_{mnrs},
\label{eq:Lpsikin}
\ee
where i) $\psi$ are spinors under the gauge group and scalars under diffs; ii) the vierbein one form
is $e^m=e^m_\mu dx^\mu$; iii) the covariant derivative $D$ contains the gauge connection one form
$\w^m_n=\w_\mu{}^m_ndx^\mu$.  In the broken phase, this action reproduces the standard fermionic
kinetic term, including gravitational interactions.

\smallskip

In this formulation, gravity is ready for enlargement of the Lorentz gauge group to a generic group
$G$, including Lorentz and other gauge interactions.  One promotes the local Lorentz index $m$ to a
larger index $M$ in a representation of $G$, while space-time and its indices $\mu$ remain
four-dimensional.  The extended vierbein $e_\mu^M$ still transforms as a one-form under standard 4D
diffs, but $M$ is enlarged.  The gauge field $\w^M_{\mu\,N}$ of the enlarged group $G$ contains both
the spin connection and standard gauge interactions.

Let us exemplify this construction in the case of $G=\SO(1,3+N)$, which preserves the metric
$\eta_{MN}=\diag\{1,-1,-1,-1,-1\ldots\}$ with $M=0,\ldots N+3$.

Notably, a vierbein VEV can be arranged again in just four directions,
\be
\bar e_\mu^M=\left\{\ba{ll} M_{pl}\,\delta _\mu^M\,,&\qquad  \text{for }0\leq M\le3\\
  0\,,&\qquad  \text{for } 4\leq M\leq N+3\ea\right.
\label{eq:VEVext}
\ee
which does a twofold job. It breaks again diffs and the 4D part of $G$ down to global simultaneous
Lorentz transformations of $\mu$ and the first four indices $M$, and in addition it leaves unbroken
a local subgroup $\SO(N)$, mixing the last $N$ directions where the VEV vanishes.  This mechanism
was used in~\cite{Nesti:2009kk}, with $\SO(11,3)$ broken in this single step to a $\SO(10)$ GUT. As
analyzed there, the correct fermionic, gauge, and gravitational lagrangians emerge after the
symmetry breaking of the $G$-invariant unified theory, for instance from a direct generalization of
(\ref{eq:Lpsikin}).

In this work, the VEV~(\ref{eq:VEVext}) is assumed. It was shown in~\cite{Nesti:2007ka,Nesti:2009kk} that it is a solution of the connection's equations of motion, while a dynamical mean field origin was proposed in~\cite{Percacci:1984ai}. An other interesting possibility is that the vierbien is realized as bilinear condensate of more fundamental fermions, see for instance~\cite{Diakonov:2011im,Vladimirov:2012vw,Volovik:2020bpn,Wetterich:2004za}.

In~\cite{Nesti:2007ka,Nesti:2009kk} it was also discussed how this unification respects the Coleman
Mandula theorem. The point is that in the broken phase the symmetry group is indeed the direct
product of an internal gauge and global Lorentz.  Conversely, in the unified phase, a background
metric is absent and the theorem does not apply.  We refer to~\cite{Percacci:1984ai,Percacci:1990wy}
for extra discussions.

\medskip

We wish here to study the action of $\U(1)_P$ and of $P$ on the background
(\ref{eq:VEVext}). Because $\U(1)_P$ contains a gauge rotation in (e.g.)\ the 1-2 and 3-4 planes and
the VEV $\bar e_\mu^M$ is nonzero in these subspaces, the continuous $\U(1)_P$ symmetry is
broken (indeed only $\SO(N)$ survives).
The discrete $P=U(\pi)$ instead has a more interesting fate.  After this $\pi$ rotation, four
internal spatial directions change sign, $e_\mu^M\to - e_\mu^M$ for $M=1,2,3,4$.  Because the VEV is
nonvanishing only in the first three directions $M=\mu=i=1,2,3$, one can write
\bea
\label{eq:Pvev}
\quad&&\bar e_\mu^{M=0,\ldots, 3} =  M_{pl}\,{\rm diag}(1,1,1,1)     \\
&&\qquad \stackrel{U(\pi)}{\to} \    M_{pl}\,{\rm diag}(1,-1,-1,-1)\,,\nn
\eea
or $\bar e_\mu^M\to \bar e_\mu^M\eta^{MM}$ (no summation). Thus also $P=U(\pi)$ is broken, as it
does not preserve $\bar e_\mu^M$.

We however notice that the VEV can be restored by adding a $I_s$ spatial inversion,
$e_\mu^M\to \eta_{\mu\mu}e_\mu^M$, which completes the action of $\P$. We then find
\be
\P:\quad \bar e_\mu^M\  \to\   \eta_{\mu\mu}\bar e_\mu^M  \eta^{MM}= \bar e_\mu^M\,,
\ee
i.e.\ the vierbein VEV is invariant under combined internal parity and spatial inversion,
$\P=I_s\circ P$.  This result shows that the breaking mechanism glues not only the gauge and diff
Lorentz transformations but also glues internal parity with spatial inversion, to produce the
standard behavior of parity in the low energy field theory.

Thus, if the Lagrangian is invariant under space inversion, then the low energy theory will be exactly $\P$ invariant.


\Sec{Emergence of LRSM Yukawa terms.}
It is instructive to explicitly discuss, in the $\SO(1,7)$ example, the emergence of the
$\P$-invariant fermionic Yukawa terms of the LRSM.  While in the symmetric phase a direct (Majorana)
mass term for the fermions $\Psi ^t C\Psi$ is forbidden by the other gauge interactions, e.g.\ $B-L$
and color (or $\SU(4)$) one can have Yukawa terms by introducing some extra bosonic field, for
instance a generic (reducible) $H\in 8_c8_c^\dag$:
\be
\label{eq:yuk}
\L_{Y\!uk}= Y_H \Psi^\dag H \Psi+h.c.\,,
\ee
where $Y_H$ is a generic complex Yukawa matrix.

\pagebreak[3]

It is also useful to spell out the decomposition of
$\Phi$ under the breaking $\SO(1,7)\to \SU(2)_L\times \SU(2)_R\times \SO(1,3)$,
{\small%
\bea
H\,&&=
\left(1_L+3_L,1_R, \overline{2_l}2_l\right)+\left(1_L,1_R+3_R,\overline{2_r}2_r \right)\nn\\
&&\quad {}+\left(2_L^*,2_R,\overline{2_l}2_r \right)+\left(2_L,2_R^*, \overline{2_r}2_l\right)\nn\\
&&=L_\mu \left(1_L,1_R,4_l\right)+ L_\mu^a \left(3_L,1_R,4_l\right) +L_i \left(2_L^*,2_R,3_{\bar l r}\right)\nn\\
&&\quad {}+ \Phi_{\bar LR} \left(2_L^*,2_R,1_{\bar l r}\right) \quad + L\leftrightarrow R\,.
\label{eq:phipieces}
\eea}%
Thus $H$ contains Lorentz 4-vector, 3-vector, and singlet representations transforming under the weak groups.

The last term represents a scalar bidoublet, as found in Left-Right symmetric theories, where its
weak scale VEV breaks electroweak symmetry and gives standard masses to fermions.  We find actually
two independent such complex bidoublets, $\Phi_{\bar L R}=\Phi_1$ and $\Phi_{\bar R
  L}=\Phi_2^\dag$. Decomposing $Y_H$ in hermitian components, $Y_H=Y+ i \tilde Y$, (\ref{eq:yuk})
becomes a generic Yukawa lagrangian
$\bar \psi_L[Y(\Phi_1+\Phi_2)+\tilde Y i(\Phi_1-\Phi_2)]\psi_R+h.c.$.  The invariance under $\P$ is
confirmed by noting the bidoublets transformation $\Phi_{1,2}\leftrightarrow \Phi_{2,1}^\dag$.

Now, since the minimal LRSM has only one bidoublet, a fact tied to the nice model predictivity, one
may be tempted to restrict the $H$ field.  However, the only possibility is to assume a hermitian
representation, $H\equiv H^\dag$, i.e.\ $\Phi_1\equiv \Phi_2$, but this would lead to unrealistic
fermion masses, given by the sole matrix $Y$.
The natural possibility is instead to allow generic $\Phi_1$, $\Phi_2$ fields and realize that after
the $G$ breaking at Planck scale, only one combination can (and has to) be kept light, with mass at
the $v_R$ scale, and identified with the LRSM bidoublet.
The situation is parallel to what happens when embedding the SM into the LRSM: the SM Higgs doublet
$\phi$ may be rewritten as a \emph{real} LR bidoublet $\Phi$($\equiv \epsilon \Phi^*\epsilon$), but
then the Yukawa lagrangian would unrealistically allow just a single hermitian matrix.  One
considers thus a \emph{complex} bidoublet: one real component is kept light and identified with the
SM Higgs doublet at weak scale by careful choice of coupling constants ($\mu$-terms); without further
choices, the other remains naturally heavy at the high ($v_R$) breaking scale.
In the present framework, one combination $\Phi$ of the two above bidoublets shall be kept light and
leads effectively to LRSM Yukawa lagrangian,
\be
\L_{Y\!uk}\to\bar \psi_L\left[Y \Phi+\tilde Y \tilde \Phi\right]\psi_R+h.c.\,.
\label{eq:LyukLRSM}
\ee
The other bidoublet has a natural mass at the Planck breaking scale, disappearing from the low
energy spectrum.  Incidentally, the same fate can be assumed for all the other components
transforming nontrivially under Lorentz in (\ref{eq:phipieces}), $L_\mu$, $L_\mu^a$, $L_i$, also
avoiding possible issues with the signature of their nonstandard kinetic terms, see discussion below.


Similarly, one can implement Majorana Yukawa terms for fermions as
$\L_{Maj}=Y_\Delta \Psi^t \Delta \Psi$ where, still in $\SO(1,7)$, $\Delta$ transforms in the
$8_s8_s$ representation. Its decomposition contains the two $SU(2)_{L,R}$ triplets
$\Delta_{L,R}$, and generates the standard Yukawa terms
$Y_\Delta\psi_L^tC\Delta_L\psi_L+L\leftrightarrow R$, leading to Majorana masses for neutrinos
via type-I and II seesaw.  As above, several field components which transform nontrivially under
Lorentz are present and naturally have mass at the Planck scale.

\smallskip
 
A comment is in order regarding the consequences of having noncompact gauge groups, which are known
not to have finite-dimensional unitary representations.  Indeed, the $L_\mu$, $L_\mu^a$ and $L_i$
states appeared above in eq.~(\ref{eq:phipieces})\ are a manifestation of this fact.  This is
potentially a serious problem, that could make the whole approach fundamentally flawed.  A possible
solution argued above is that no ghost state shall have mass below the Planck scale.  A complete
modeling, going beyond the scope of this study, should pay special attention to this requirement.
It is worth recalling that states with Planck mass and seemingly negative kinetic terms appear also
in generic gravitational theories with propagating torsion, regardless of
unification~\cite{Sezgin:1979zf}.  Various proposals to circumvent this problem exist in the
literature (see e.g.~\cite{Julve:1978xn,Bender:2007wu}) including recent ones, where the possible
metastability of ghost states is investigated~\cite{Gross:2020tph}, or their quantization with a
dedicated prescription is proposed~\cite{Anselmi:2020opi}.

We can add on top of these possibilities, that the standard tree level analysis is hardly
conclusive, as these ghosts occur with the transition to a different regime.  Indeed, in the
symmetric phase, as noted in~\cite{Percacci:1984ai,Percacci:1990wy,Nesti:2009kk}, the theory does
not possess a background metric and has no quadratic kinetic terms. It thus belongs to a topological
nonperturbative phase of quantum gravity, where new representations may appear as bound states.
Interestingly, recent proposals where the vierbein is built as from fermion biliners, may help in
dealing with these issues, see e.g.~\cite{Wetterich:2004za, Wetterich:2021cyp}
and~\cite{Diakonov:2011im,Vladimirov:2012vw,Obukhov:2012je}.

These comments apply to the $\SO(1,7)$ example and to the more general groups that we discuss
now.


\Sec{Complete unifications and other symmetries.} The analysis of other groups and the inclusion of
strong interactions can proceed straightforwardly: a good path is to pre-unify color SU(3)\ and
hypercharge U(1)\ into $\SU(4)\approx\SO(6)$ of Pati and Salam~\cite{Pati:1974yy}, ready to be
included in a pseudo-orthogonal group. Considering in generality $\SO(p,q)$, we display in
Table~\ref{tab:groups} the realistic cases involving weak interactions, which we briefly comment.

First, from the $\SO(1,7)$ example above, we have seen that the $R_i$ generator involved in $P$ is a
cross rotation between one spatial Lorentz direction and one relative to $\SO(0,4)$. It is then
clear that if $\SO(4)$ were to be included with a time-like signature, like $\SO(4,0)$ inside
$\SO(5,3)$, then $\P$ could not be achieved, as the cross generators are noncompact,
boost-like. Instead, one can rotate one of the $\SO(4)$ directions with internal direction 0 to obtain
its inversion, and the VEV may be preserved by adding a time inversion $I_t$.  We indicate the
symmetry as $\T$ in the table: it amounts to time-reversal plus exchange of the Left and Right weak
groups. Its enforcement leads to \emph{real} Yukawa matrices, thus requiring spontaneous CP
violation, which is not so appealing at least from the point of view of model minimality.

\pagebreak[3]
\begin{table}
  \def\ot{\!\!\!\mathop{\raisebox{.3ex}{$\scriptstyle\otimes$}}}
  \small
  \def\arraystretch{1.1}
  \centerline{$\begin{array}{|c|rrrrr|ccc|}
    \hline
  \text{\bf   p+q = 8 }& \rlap{\bf spinor\,=\,{\bf 16$_\RR$}\,(Majorana)}&&&&&&&\\[.7ex]
\SO(1, 7)  & &        &\SO(1, 3) &\ot \SO(0,4) &   &   &   & \P    \\
\SO(5, 3)    & &\SO(4, 0)   &\ot\SO(1, 3) &       &   &  & \T  &   \\
    \hline
\text{\bf  p+q\,=\,14 } & \rlap{\bf spinor\,=\,{\bf 64$_\RR$}\,(Majorana-Weyl)} &&&&&&&\\[.7ex]
\SO(7, 7)  & & \SO(6, 0) &\ot\SO(1, 3)&\ot \SO(0, 4)       &    &   & \T_{col} , &\P \\
\SO(11, 3)  & &\ \SO(10, 0)   &\ot \SO(1, 3)   &   &       & \C\   &  \T_{so10} &  \\
    \hline
    \end{array}$}%
  \vspace*{-0.5ex}
  \caption{Breaking of unifying orthogonal groups and emerging discrete symmetries.
    \vspace*{-2ex}
  }
  \label{tab:groups}
\end{table}

Complete unifications involving strong interactions can give rise to more general discrete
symmetries. We list in the table the realistic cases, which can be implemented only by the groups
$\SO(11,3)$ (proposed in~\cite{Nesti:2009kk,Krasnov:2021jjp}) or $\SO(7,7)$ (also proposed
in~\cite{Maraner:2003sq,Krasnov:2018txw}).  In both cases, the minimal Majorana-Weyl spinor representation has real
dimension $64$ which, when mapped into 32 complex~\cite{Nesti:2009kk}, leads precisely to a complete
SM family,
\be
\mathbf{64}_\RR\equiv\mathbf{32}\to(\mathbf{2}_l,\mathbf{2}_L, \mathbf{1}_R, \mathbf{4})\oplus (\mathbf{2}_r,\mathbf{1}_L,
\mathbf{2}_R, \mathbf{4})\,,
\label{eq:64PS}
\ee
in Lorentz$\,\times\,$Pati-Salam notation.

Other groups as $\SO(1,13)$ or $\SO(5,9)$ are not viable
as they have only \emph{symplectic} Majorana-Weyl representations, leading to extra mirror families of
opposite chirality.

In the $\SO(7,7)$ case, $\SO(4)$ is present with spatial signature and leads to $\P$, but $\SO(6)$
is included as time-like.  By an analysis similar to above, one finds a new discrete symmetry,
amounting to time-reversal plus SU(4) color conjugation, named $\T_{col}$ in the table.  This
additional discrete symmetry may or may not survive the lower stages of symmetry
breaking.

We stress that the breaking of $\SO(7,7)$ has arguably to proceed in one step at Planck scale, at least to
the Pati-Salam subgroup $\SO(4)\times\SO(6)$, so that the noncompact generators of $\SO(4,6)$ have
mass at or above the Planck scale. 

In the case $\SO(11,3)$ we find an analogous symmetry, $\T_{so10}$, while $\P$ is
absent. $\T_{so10}$ may lead to $\T$ and/or $\T_{col}$, depending on the $\SO(10)$ breaking pattern.
In the table, we list also the more standard $\C$ LR-symmetry, i.e.\ charge conjugation plus exchange
of Left and Right weak groups, which is part of $\SO(10)$.

\smallskip

We confirm that, in the Pati-Salam notation (\ref{eq:64PS}), $\P$ acts linearly by exchanging the
two Left and Right components, while $\T$ and $\T_{so10}$ act antilinearly, replacing the spinor
with its complex conjugate, as is required for a time-reversal (see also~\cite{Nesti:2009kk} for a
discussion of antilinearity of broken generators).

\smallskip

\Sec{Discussion on Parity and Strong CP.} We
have shown that by considering the presence of a high scale gauge symmetry unifying local Lorentz
and gauge interactions, the theory automatically enjoys $\P$-parity symmetry below the first stage
of symmetry breaking. This motivates the framework of Left-Right symmetric theories with $\P$ as
exact LR symmetry (then broken spontaneously at a lower scale).

However, it is necessary to deepen and clarify our understanding of this result.  We established
that $\P$ arises from the gluing of internal parity $P$ and spatial inversion $I_s$. This leads us
to consider the possibility that the theory respects $P$, as a gauge symmetry, but might still
violate space reflection. An example is the analog of the QCD theta term, namely
$\theta F^M_N\wedge F^N_M$, the two-form $F^M_N$ being the curvature of the connection one-form
$\w^M_N$. This term respects internal parity $P$ because it is gauge-invariant but violates space
inversion.  As a result, in the low-energy theory, it leads to a term such as
$\theta F\tilde F$, which violates $\P$.\footnote{The $\P$-violating term $\t F\tilde F$ may be rotated away in the quark masses via the anomaly~\cite{Fujikawa:1979ay}. The Yukawa couplings in
  (\ref{eq:LyukLRSM}) become non-hermitian and only preserve a new internal parity $P'$, because the
  chiral rotation does not commute with $U(1)_P$. The model is still $\P$-violating, but the
  non-hermiticity in (\ref{eq:LyukLRSM}) lies in an overall phase $\t$ at most.}

The spatial inversion could thus be assumed or not to be an invariance of the theory. This is in
accord with the fact that, while in 4 dimensions diffeomorphisms have two disconnected components,
the proper one and the one including a reflection, General Relativity is formulated as invariance
under the proper component only. One might thus assume invariance under the full diffeomorphisms as
a funding principle, and the theory would have no $\P$ violating terms.  This choice can be viewed
as a solution to the Strong CP problem, as
in~\cite{Senjanovic:1975rk,Senjanovic:1978ev,Mohapatra:1979ia}, see~\cite{Senjanovic:2020int}.

In a more physical approach, one shall test this hypothesis by considering possible violations of
spatial inversion in the theory.  In the present context, the unification of the internal symmetries
leads at least to the prediction that various $\P$-violating terms, regarding different
interactions, will be connected.

For instance, one of the most stringent tests is the experimental bound from the electric dipole
moments (e.g.\ of the neutron~\cite{Abel:2020gbr}). The relative bounds of the order
$\bar \theta<10^{-10}$ directly translate for us into limits on the gravitational analogous,
$\theta\tilde RR$. This is argued to be physical~\cite{Deser:1980kc}, and the question of how
it could be measured is the subject of some recent studies,
e.g.~\cite{Dvali:2016uhn,Arunasalam:2018eaz,Chen:2021jcb}.

On the other hand, the bounds on parity-violating Chern-Simons extensions~\cite{Yunes:2009ch} would
be connected with the QCD axion terms.
Another example breaking spatial parity but not the gauge symmetry is the Immirzi term $\alpha R^{MN}\wedge e_M\wedge e_N$, although there is practically no bound on it from semiclassical effects~\cite{Freidel:2005sn}.
More in general, a detailed program investigating all possible terms violating spatial parity could be undertaken, along the
lines of the analysis for standard Cartan gravity~\cite{Baekler:2011jt}.


\Sec{Phenomenological implications.}  In the LRSM, the discrete parity $\P$, among many constraints on the parameters of the model~\cite{Maiezza:2010ic,Bertolini:2012pu,Maiezza:2016bzp,Maiezza:2016ybz,Bertolini:2019out}, imposes that the QCD $\theta$ is strictly zero~\cite{Mohapatra:1978fy}.  In this case $\bar{\theta}$ is computable, and nEDM together with other CP-violating observables was shown to put strong bounds on the right-handed scale~\cite{Maiezza:2014ala,
  Senjanovic:2020int,Ramsey-Musolf:2020ndm,Dekens:2021bro,deVries:2021sxz}. The RH scale is pushed beyond $\sim 28$\,TeV~\cite{Maiezza:2014ala, Bertolini:2019out}.

The present framework instead motivates also the situation as described in Ref.~\cite{Maiezza:2010ic,Bertolini:2019out}, namely, $\bar\t$ is free. In this case, $\P$ symmetry is valid in the Yukawa sector but strong CP poses no additional constraints, in complete analogy with the case of $\C$ symmetry~\cite{Maiezza:2010ic}. In this scenario, the $W_R$ scale can be lowered to $\sim 6$\,TeV.

The future LHC runs and next-generation colliders would be able to probe $W_R$ up to $\sim 30$\,TeV~\cite{Nemevsek:2018bbt}, and the potential discovery of $W_R$ in this range would point to the second scenario, where $\theta$ is nonzero and determined.  In this case, the striking consequence is that together with the validity of $\P$ in the quark sector~\cite{Maiezza:2010ic,Senjanovic:2014pva}, one would test predicted correlations between the various electric dipoles of neutron and nuclei, as analyzed in~\cite{Ramsey-Musolf:2020ndm}.  This would help to clarify the underlying mechanism behind~$\mathcal{P}$.

\Sec{Cosmic strings.}
It is noteworthy that, although the present framework contains the breaking of a continuous symmetry
to a discrete one, cosmic strings~\cite{Kibble:1982dd} do not appear. This can be understood by
looking at a possible transformation of the vierbein VEV along a closed path around a string: with
the gradual 3-4 plus 1-2 rotation up to final angle $\pi$, the result is the new
VEV~(\ref{eq:Pvev}).  This would be matched with the starting one by inverting the spatial
coordinates $\mu=1,2,3$ as mentioned above. However, inverting them just on the final part of the
path is not legitimate in a given space-time configuration, because the whole space would be
nonorientable, a situation that clearly can not be generated by standard physics like gravitational
collapse or phase transition. In practice, asking for space-time orientability rules out the
possibility of a nontrivial $\P$ around a string.

It is interesting to speculate whether one may semiclassically create such nonorientable cosmic
strings \emph{in pairs}, on the line of nonorientable gravitational
instantons~\cite{Chamblin:1996qs}. Traveling around one such string one would be faced with the
$\P$-symmetric physics. In any case, the spontaneous breaking of $\P$ at a
lower scale would attach domain walls to these strings, which would preclude the view of space
nonorientability, quite an exotic situation. Analogous comments apply to the emergent $\T$ symmetry.

\smallskip
\smallskip

A final word is worth about topological defects that may arise at the lower scales of symmetry
breaking, such as domain walls or GUT monopoles~\cite{Vilenkin:1984ib,Chakrabortty:2020otp}. In our
framework, their appearance can not be cured by nonrenormalizable operators from gravity as
in~\cite{Rai:1992xw}, but other ways out include low scale inflation or symmetry nonrestoration at
high temperature~\cite{Mohapatra:1979bt,Weinberg:1974hy}.

\pagebreak[3]

\Sec{Summary and outlook.}  We have proposed a framework for UV completion of $\P$-parity, and thus
of the LRSM in its original formulation, where $\P$ was introduced as the LR restoration of standard
parity.

We have first decomposed $\P=P\circ I_s$ into simple space inversion $I_s$ plus internal LR and
chirality exchange $P$. Then, we have shown that $P$ can be made continuous and gauged by embedding
it into the $\SO(1,7)$ or $\SO(7,7)$ gauge groups, unifying Lorentz with gauge interactions and
pointing respectively to gravi-weak and gravi-GUT models.  In these scenarios, diffeomorphism and
gauge symmetry are disentangled at high energy and are broken together at the Planck scale in such a way
that the standard \emph{global} Lorentz symmetry remains. The breaking also preserves the
simultaneous $P$ and $I_s$ transformations. This guarantees invariance under $\P$, which is thus
also protected by the gauging of $P$.

\smallskip

Space coordinate reflection $I_s$ needed an additional discussion.  Because it is not strictly
included in (proper) diffeomorphisms, one can choose whether to assume it as an additional
fundamental invariance or not.

By enlarging diffeomorphisms with $I_s$, the low-energy theory is automatically $\P$ invariant.  In this case, a direct implication is that there are no non-renormalizable $\P$-violating operators
from gravity.

In case basic spatial inversion symmetry is not assumed, one still deals with internal $\SO(1,7)$ or
$\SO(7,7)$ gauge groups, leading to a mostly $\P$-invariant low energy theory, save for a few
$\P$-violating terms that can now appear. One notable case is a topological term $\t F \tilde{F}$ in
the QCD lagrangian -- unified with the gravitational equivalent $\t R \tilde{R}$.  Therefore, even
if the theory enjoys $\P$ symmetry in the quark and scalar sectors, it does not require a vanishing
of $\bar\theta$.  In other words, protecting $P$ by gauge invariance alone does not solve the strong CP
problem.

On the other hand, this scenario has direct links with the phenomenology of the LRSM.  In that context, exact $\P$ symmetry is at the basis of predictivity in the flavor sector but was also used to attack the strong CP problem, where the nEDM limit implies a strong lower bound on the $W_R$ mass.  As we discussed, the possibility of nonzero $\bar\t$ motivates the scenario of lower $W_R$ accessible at forthcoming colliders.

\smallskip

The choice of various unifying groups has uncovered the possibility of novel low energy discrete
symmetries, such as time-reversal $\T$ involving LR exchange or $\T_{col}$ involving color
conjugation. Their analysis is left for future work.

A further property of our framework is that, although featuring a transition from continuous to a
discrete symmetry, due to the role of space there is no appearance of cosmic strings, avoiding
related cosmological issues.

Summarizing, in this work we have established a proof of concept for the gauge protection of $\P$ as
the remnant of a high energy unified gauge group and investigated the relative symmetry structure
and breaking mechanism.  After this stage, a viable model unifying gravitational and BSM degrees of
freedom will be the next outstanding challenge. We can speculate that the large symmetry structure
will pose stringent constraints on the unified model.

\smallskip

\Sec{Acknowledgements.}
The work of F.N.~was partially supported by the research grant No.~2017X7X85K
under the program PRIN 2017 funded by the Ministero dell'Istruzione,
Universit\`a e della Ricerca (MIUR).

\bibliographystyle{utphysmod}
\bibliography{PGTbib_v07}

\end{document}